\documentclass[prb,preprint]{revtex4-1} 

\usepackage{amsmath}  
\usepackage{amsfonts} 
\usepackage{graphicx} 
\newcommand{\dg}{\mbox{$^{\circ}$}}
\newcommand{\ket}[1]{\mbox{$|{ #1}\rangle$}}
\newcommand{\bra}[1]{\mbox{$\langle { #1}|$}}

\begin{document}


\title{A Demonstration of Quantum Key Distribution with Entangled Photons for the Undergraduate Laboratory}

\author{Aayam Bista, Baibhav Sharma and Enrique J. Galvez}
\email{egalvez@colgate.edu} 
\affiliation{Department of Physics and Astronomy, Colgate University, Hamilton, NY 13346}



\date{\today}

\begin{abstract}
Now that fundamental quantum principles of indeterminacy and measurement have become the basis of new technologies that provide secrecy between two communicating parties, there is a need to provide teaching laboratories that illustrate how these technologies work.  In this article we describe a laboratory exercise in which students perform quantum key distribution with single photons, and see how the secrecy of the communication is ensured by the principles of quantum superposition and state projection. We used a table-top apparatus, similar to those used in correlated-photon undergraduate laboratories, to implement the Bennett-Brassard-84 protocol with polarization-entangled photons. Our experiment shows how the communication between two parties is disrupted by an eavesdropper.  We use a simple quartz plate to mimic how an eavesdropper intercepts, measures, and resends the photons used in the communication, and we analyze the state of the light to show how the eavesdropper changes it.

\end{abstract}

\maketitle 

\section{Introduction} 

Quantum entanglement is a fundamental property of quantum systems that has recently moved from being a scientific curiosity to becoming central to modern and emerging applications. 
It is the basis of many phenomena investigated under the umbrella name of quantum information, where quantum states of light and matter are used to encode and process information.\cite{StauchAJP16,Nielsen} This field has been partly stimulated by technological advances, but mainly by a renewed appreciation of quantum physics. This appreciation has led to the teaching of quantum mechanics with textbooks that emphasize linear algebra more than wave mechanics, and with undergraduate laboratory experiments that directly illustrate the fundamentals of quantum mechanics.\cite{GalAJP14}

By enabling applications, quantum technologies have become relevant to fields of engineering and computer science as well as physics and chemistry. Thus, the academic community needs to create curricula to educate an emerging workforce. One of the first applications to arise in quantum information uses the fundamentals of quantum physics to ensure the secrecy of communication by generating a secure cryptographic key with which to encrypt messages. The advantage of the technique is  that it can reveal the presence of an eavesdropper if the communication is compromised. This technique is known as quantum key distribution (QKD), and interest in it has been stimulated by work on the development of a quantum computer. We expect that a quantum computer will be able easily and quickly to decode the cryptographic keys commonly used today.  QKD offers a defense against such code breaking. Cryptographic keys generated using quantum physics will provide immunity against eavesdropping and also security against future decryption because some future quantum computer will not be able to decode past secret communications. The development of this technology has made significant progress and is already being offered commercially as an option in secret communications.  

At a fundamental level, QKD is based on the transmission of a single photon in a quantum state by a sender, and in the measurement of this state by a receiver. In its most simple form, the apparatus can have two settings: in one setting the photon is in an eigenstate state of the apparatus and is measured with  absolute certainty; in the other setting the photon is in a superposition of eigenstates, and the result of the measurement is indeterminate. The same would be true of an eavesdropper trying to extract the information carried by the photon. 
If the eavesdropper intercepts the photon in a state of superposition, no useful information will be extracted form it, and in resending the photon to avoid revealing the intrusion, the photon will no longer be in the initial state. Instead it will be in a state modified by the act of measurement by the eavesdropper. 
This modification of the state of the photon can be used by the sender and receiver to recognize the presence of an eavesdropper. 
The experiment also demonstrates the non-classical essence of quantum physics, succinctly put by P.A.M. Dirac: \cite{Dirac} when a photon in a superposition of two states is measured, the photon ``has to make a sudden jump from being partly in each of these two states to being in entirely one or the other of them.'' 
More generally, the act of measurement and gaining information from it produces a disturbance,\cite{FuchsPeresPRA96} and the disturbance can be detected.

The teaching of this technological application of quantum physics is already facilitated by an excellent textbook,\cite{LoeppWooters} and a more general tutorial on the subject in a recent book,\cite{Raymer} but more explanations can be found online (searching for QKD). Instructional research on this technique has received recent attention.\cite{DeVorePERC14,KohnleEJP17}  
Teaching-laboratory demonstrations that use common light sources (e.g. lasers, LED's) either home-made\cite{Utamaarxvii19} or commercial\cite{Thorlabs} involve many photons. They provide an illustration of the principle but do not demonstrate the phenomenon with single quanta. We have developed and implemented a table-top experiment that uses single photons to illustrate and demonstrate the fundamental principles behind this application, and we are adapting it into a curricular offering. To our knowledge this type of laboratory exercise has not been presented before.

In this article we describe our development of this laboratory experiment that shows how quantum physics allows one to use entangled photons to produce an encryption key that is knowable only to the sender and receiver. Should an eavesdropper intercept the communication, quantum features will make the intrusion apparent. 
Section~\ref{sec:principles} explains the fundamental principles.  Section~\ref{sec:imp} presents the laboratory technique. Section~\ref{sec:results} presents our results. Section~\ref{sec:qst}  presents the results of using the technique of quantum state tomography to measure the state of the light within the context of QKD. Section~\ref{sec:concl} presents our conclusions. There are two appendices: Appendix~A has a brief description of the encryption of the communication once the key is obtained; and Appendix~B has information about implementing quantum state tomography.

\section{The Underlying Quantum Principle} \label{sec:principles}

A basic postulate of quantum mechanics is that a device that measures a physical quantity on a system is represented by an operator. The outcome of a measurement is an eigenvalue of the operator. After the measurement, the system is left in the eigenstate corresponding to the eigenvalue.\cite{Sakurai} If the initial state of the system is one of the eigenstates of the measurement apparatus, then a measurement will leave the system 
in the same state with unit probability; but when the initial state is not an eigenstate, then the act of measurement projects the initial state onto one of the eigenstates of the measuring device with a probability equal to the absolute value squared of the inner product of the initial and final states. Once the state is projected, it has no memory of the initial state.  A consequence of this postulate is that when we do a measurement we are likely to change the state of the system in an irrevocable way. The probabilistic outcome further embodies another aspect of quantum physics: the outcome of an act of measurement is inherently random weighted by the corresponding probability. Thus, there is no way to know beforehand with full certainty which eigenstate will be the final state.

A second important quantum principle for QKD 
is the no-cloning theorem: the state of a system cannot be cloned.\cite{WootersNat82} If our system is, for example, a photon in an arbitrary state, we cannot duplicate the state without destroying the initial state. This theorem has easily accessible analytical demonstrations.\cite{Townsend,Beck} Interestingly, it has been demonstrated that we can destroy the initial state of the system and recreate it in a remote location without physically sending it, a process known as quantum teleportation.\cite{BennettPRL93,BouwmeesterNat97} This finding is also easily described analytically within the formalism of introductory quantum mechanics.\cite{Townsend,Beck,McIntyre}

The two aspects of quantum theory outlined above can be used to achieve secure communication: If we send a single photon in a quantum state, then if an eavesdropper intercepts the photon to measure its state and resend it (known as intercept-resend action), 
it will likely modify the state of the photon and introduce errors that will reveal that the communication has been compromised. The sender and receiver can use state superposition to exchange information. In doing so they need to sacrifice part of the communication to identify a possible intrusion. This quantum exchange is not practical to do with the actual communication, so it is instead done with the key generated to encrypt the message (see Appendix A for an example). 

The use of a one-time key, generated with each communication, ensures security.\cite{GisinRMP02,ScaraniRMP09} 
Quantum key distribution (QKD) is thus a method of producing a secret key based on quantum principles. This is a topic of current research, and efforts to improve its implementation and error correction are ongoing. These efforts include developing various types of encoding besides polarization (discussed here), such as energy-time entanglement, distributed phase and continuous variables.\cite{ScaraniRMP09,LoNP14} Novel directions in this technology include the potential use of high-dimensional states.\cite{BouchardQ18} Currently the challenges of single-photon sources (efficiency and cost) are overcome by the use of attenuated beams (with more than one photon per pulse), but encoded with intensity variations  and selective use of pulses in a way that mitigates the eavesdropping done by splitting the pulse,  a method called decoy-state.\cite{WangPRL05,LoPRL05}  This technology is already in place and commercially available by vendors such as IDQuantique, MagiQ and Quantum XChange. However, research on single-photon QKD continues and may become a reality in the future with technological advances.\cite{KupkoNP20}

\subsection{BB84 protocol}\label{sec:bb84}

The most basic and effective technique for quantum encryption was invented by Bennett and Brassard in 1984, and is known as the BB84 protocol.\cite{Bennett84} It uses the polarization of the light, a 2-state system, consisting of horizontal and vertical polarization states (picking a simple coordinate system), and one of its two mutually unbiased bases (MUB), with diagonal ($+45\dg$ from horizontal) and antidiagonal ($-45\dg$ from horizontal) basis states. In MUB's, all eigenstates of one basis are measured with equal probability by the states of the other basis. This is a critical component of QKD: If we send information in the form of a state in one basis, then a measurement in another MUB will yield no information about the initial state.

We can put this in a more analytical form. If our basis consists of the states of horizontal and vertical polarization $\ket{H}$ and $\ket{V}$, respectively, then the (MUB)  diagonal basis states are 
\begin{eqnarray}
\ket{D}&=&\frac{1}{\sqrt{2}}\left(\ket{H}+\ket{V}\right),\label{eq:D}\\ 
\ket{A}&=&\frac{1}{\sqrt{2}}\left(\ket{H}-\ket{V}\right).\label{eq:A}
\end{eqnarray}
Measuring a photon initially in state $\ket{D}$ using the horizontal-vertical ($HV$) basis,  projects states $\ket{D}$ and $\ket{A}$ onto state $\ket{H}$ with probability 
\begin{equation}
P_H=\left|\bra{H}D\rangle\right|^2=\left|\bra{H}A\rangle\right|^2=1/2,
\end{equation}
and similarly, onto state $\ket{V}$ with probability 
\begin{equation}
P_V=\left|\bra{V}D\rangle\right|^2=\left|\bra{V}A\rangle\right|^2=1/2.
\end{equation}
Thus, measurement of the photon in state $\ket{D}$ or $\ket{A}$ yields no information about the state when measured in the $HV$ basis.
Similarly, a photon initially in state $\ket{H}$ or $\ket{V}$ measured using the diagonal-antidiagonal ($DA$) basis also yields no information about the initial state of the photon because it gets projected to states $\ket{D}$ or $\ket{A}$ with equal probability. However, if we measure the state of the photon (e.g. $\ket{H}$) in the basis where its state is one of the eigenstates (i.e., $HV$ basis), then we obtain the state of the photon ($\ket{H}$) with unit probability.

The idea  for the protocol is then that the sender, normally called Alice, sends a photon in an eigenstate of a given basis  to the receiver, normally called Bob. The basis in which Alice sends her photon is generated randomly.  Bob also picks the detection basis randomly. For this, both agree on a convention on the information, such as, for example, $\ket{H}$ and $\ket{D}$ states constituting a ``1,'' and $\ket{V}$  and $\ket{A}$ constituting a ``0''. When Alice and Bob use the same basis, then Bob will get the state of Alice's photon; but when they use different bases, then there is probability of 1/2 that Bob will not get the state sent by Alice. 
Because of this, after completion of the communication, Alice and Bob must compare the bases that they used (but not the results of the measurements), and throw away the data that was obtained when they used different bases. This communication can be made over a public channel because it does not reveal the actual outcome of the measurements.
Bob will get Alice's state when he uses the same basis as Alice, and so they only keep those data (i.e., a string of 1's and 0's), which becomes the elements of the key.  (The key is then used in an encrypted communication, as described in Appendix A.)

The single-photon protocol described above cannot be used to send information securely with many photons.
If she sends a beam, consisting of many photons in the same state, then the eavesdropper can use a beam splitter to get a sample of the photons and then measure them without disturbing the communication. If only one photon is used in the communication, then an eavesdropper will have to measure it. She will have to pick a basis, and if she does not pick the same basis as Alice, then she will introduce detectable errors in the communication due to her modification of the state of the photon. 
A convenient source of single photons uses spontaneous parametric down-conversion to produce photon pairs to later detect them at coincident arrival times.\cite{GalAJP14} Attenuated beams are not a good source of single photons because there is a finite probability of more than one photon traveling at any given time.\cite{GalAJP14}

A third element of the communication between Alice and Bob involves the random choice of Alice's initial state (i.e., for sending a random string of 1's and 0's). 
The quantum physics of entangled photons already has an inherent randomness that can be used for this purpose.\cite{EkertPRL91} If we use entangled states of 2 photons that Alice and Bob can share, such as the state
\begin{equation}
\ket{\psi}=\frac{1}{\sqrt{2}}\left(\ket{H}_1\ket{H}_2+\ket{V}_1\ket{V}_2\right),
\label{eq:psi+}
\end{equation}
where $1$ and $2$ stand for Alice and Bob's photon, respectively, then a measurement in the $HV$ basis has an intrinsic probability of 1/2 of resulting in a 1 (i.e., measuring $\ket{H}$) or a 0 (i.e., measuring $\ket{V}$). For example, if Alice creates a pair of entangled photons in the state given by Eq.~\ref{eq:psi+}, keeps one photon and sends the other one to Bob, then if Alice performs a measurement on her photon in the $HV$ basis, she will obtain $\ket{H}_1$ or $\ket{V}_1$ with 1/2 probability in each case. In doing so, she also projects the state of the photon heading toward Bob to $\ket{H}_2$ or $\ket{V}_2$, respectively. (This is a useful way to think of it although quantum mechanics predicts the correlations but not causal actions.) Thus, because of quantum indeterminacy, it is truly random whether Alice and Bob share a 1 or a 0.  
More strictly, Alice's measurement is a projection of the state, which in turn defines the state of the photon received by Bob. If the outcome of her measurement is horizontal, then the projection becomes
\begin{equation}
\ket{H}_1\bra{H}_1\;\psi\rangle=\frac{1}{\sqrt{2}}\ket{H}_1\ket{H}_2,
\end{equation}
and similar for the other possible outcome
\begin{equation}
\ket{V}_1\bra{V}_1\;\psi\rangle=\frac{1}{\sqrt{2}}\ket{V}_1\ket{V}_2.
\end{equation}
In both cases the probability is the square of the probability amplitude, the factor multiplying the product states listed above.

An excellent student exercise is to use Eqs.~\ref{eq:D} and \ref{eq:A} to show that the entangled state given by Eq.~\ref{eq:psi+} can be transformed to
the $DA$ basis, yielding 
\begin{equation}
\ket{\psi}=\frac{1}{\sqrt{2}}\left(\ket{D}_1\ket{D}_2+\ket{A}_1\ket{A}_2\right),
\label{eq:psi+DA}
\end{equation}
also a superposition of the two possibilities where both photons are parallel to each other. This means that when Alice creates the entangled state she can measure it in either basis, and if Bob uses the same basis, he will be obtaining the same state Alice obtained, and therefore the same data. Should they use different bases, they will incur in the same type of errors described above. We note that if for some reason Alice were to delay the measurement of her photon, then the state-defining projection might be made by Bob or even Eve! The results are the same regardless of the order in which the measurements are done.

\begin{figure}[h!]
\centering
\includegraphics[width=5in]{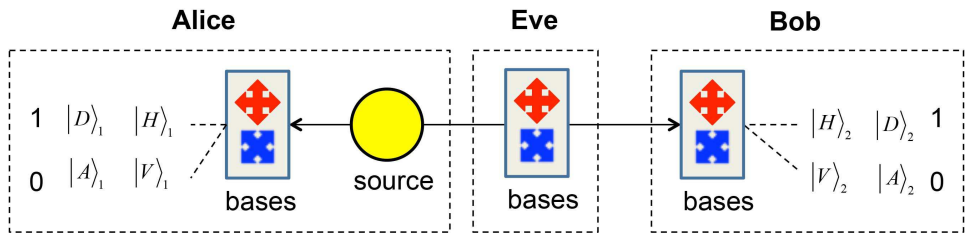}
\caption{Schematic of the BB84 protocol applied to entangled photons.\cite{EkertPRL91} The sender, Alice, generates photon pairs in the polarization-entangled state of Eq.~\ref{eq:psi+}, keeps one photon and proceeds to measure it in one of two possible bases: $HV$ or $DA$. The other photon goes to Bob, who does the same, although Alice and Bob choose their bases randomly. Between Alice and Bob is Eve, the eavesdropper, who measures the state of the photon going to Bob guessing one of the two bases and resending the state that she measured.}
\label{fig:schem}
\end{figure}

In our demonstration, shown schematically in Fig.~\ref{fig:schem}, we use the process of spontaneous parametric down-conversion to create pairs of polarization-entangled photons in the state of Eq.~\ref{eq:psi+} (or equivalently, Eq.~\ref{eq:psi+DA}). In an implementation, this would be done on Alice's end, with Alice keeping one of the photons and sending the other one to Bob. Both would use a method of separating spatially the photons by the state in which they are measured. We used polarizing splitters for both, so that when the photon reaches the appropriate detector it can be counted as 1 or 0.

\subsection{Eve}
An important  step in this demonstration is to implement Eve. 
What does Eve do? As Eve intercepts the photon she needs to pick a basis: $HV$ or $DA$. Once she picks it, she makes a measurement with a polarizing splitter and 2 detectors. Thereafter she resends the photon to Bob in the same state that she detected.  

There are two cases depending on Eve's choice of basis:
\begin{enumerate}
\item{\em Case I: Eve picks the same basis as Alice and Bob}. 
If Eve picks the same basis as Alice, then she measures the same state as Alice. She resends it to Bob, and if Bob measures it in the same basis, then he will get the same outcome as Eve, and consequently the same as Alice. In this case, Eve's interception was successful: she measured Bob's photon without him being able to know that she intercepted it. 

\item{\em Case II: Eve picks a different basis than Alice and Bob}. Let us now consider the other possibility: Alice measures, for example, in the $HV$ basis and gets state $\ket{H}$ as outcome, but Eve picks the other  basis ($DA$). Eve does not know that it is the wrong basis, and so she gets an outcome (state $\ket{D}$ or $\ket{A}$ with probability 1/2) and resends the photon in that state to Bob. Subsequently, Bob makes a measurement. If Bob does not pick the same basis as Alice ($HV$), then the outcome is immaterial because the data would be discarded anyway, as mentioned earlier. If Bob picks the same basis ($HV$) he expects the correct outcome, but because Eve measured in the wrong basis, she resends the outcome that she got ($\ket{D}$ or $\ket{A}$). As a consequence, Bob has only 1/2 probability of getting the correct outcome  ($\ket{H}$) in his basis.
\end{enumerate}

As an example of case II, suppose an entangled pair is produced and in Alice's measurement the photon goes to the detector selecting state $\ket{H}$. She records this as a 1. On its way toward Bob, Eve intercepts the partner photon, and measures its state. Eve picks the $DA$ basis, and in her measurement she obtains $\ket{D}$ (a 1). Bob then gets the photon from Eve and decides to use the $HV$ basis. Bob measures the state of the photon and obtains $\ket{V}$ (a 0). Once all the data are recorded, Alice and Bob compare the bases that they used, and keep this data point because both used the same basis ($HV$). However if they were to compare the actual results of the measurements, they would disagree, because of Eve's intervention. There is another way Bob could get this result.  If Eve would have gotten $\ket{A}$ as a result of her measurement Bob would still have a 50\% chance of measuring $\ket{V}$ for the state of the photon.

In implementing Eve's action, our apparatus should mimic the two possibilities presented above.  It would be best to set up the intercept-resend action by Eve, but that is not possible for practical reasons: because it involves absorbing a down-converted photon and the re-emitting it, which is not easily done in our demonstration. The alternative is to mimic Eve, not by absorbing the photon and re-emitting it, but by modifying it in such a way as to destroy the entanglement when Eve choses the wrong basis (i.e., different than Alice and Bob), and preserving it when she chooses the right basis (i.e., same as Alice and Bob). There is a transmissive optical method that can do this, described below, so that our implementation of Eve fulfills its role in our demonstration. 

The final step is to record single-photon events at Alice's and Bob's ends and generate a key. For completeness, in Appendix A we describe how Alice and Bob communicate once they have the key. 

\section{Implementation}\label{sec:imp}

\begin{figure}[h!]
\centering
\includegraphics[width=5in]{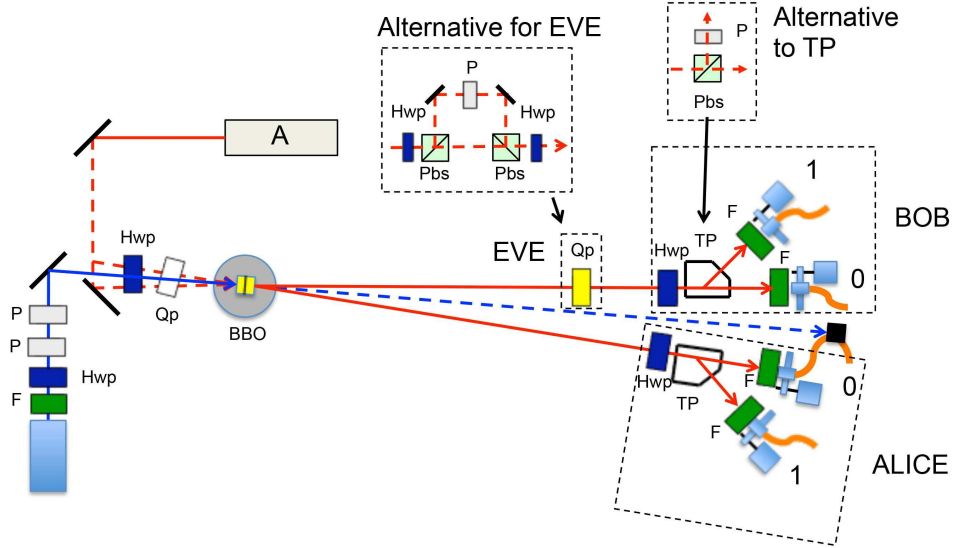}
\caption{Apparatus for implementing QKD with entangled photons (see text). Optical components include: HWPs (Hwp), polarizer (P), quartz plate (Qp), beta-barium borate crystal (BBO), alignment laser (A) Thompson prism (TP, polarizing beam splitter (Pbs), and band-pass filters (F).}
\label{fig:app}
\end{figure}

The apparatus used for this demonstration is shown in Fig.~\ref{fig:app}. We used a source of polarization-entangled photons. This setup has been described extensively before.\cite{KwiatPRA99,DehlingerAJP02} Briefly, we use a gallium-nitride diode laser operating with only current control (prices start at about \$20 for the cheapest laser ``pointers,'' but more reliable ones are available commercially for a few hundred dollars). This laser had a wavelength of 402.5 nm and output of about 50 mW. A bandpass filter removed infra-red light from the laser. For the initial alignment, the full power was used, but for the experiments we attenuated the beam by polarization projection using a half-wave plate and 2 polarizers in series (to increase the extinction ratio in the orthogonal direction), with the final  polarization aligned horizontal.  After a steering mirror (see Fig.~\ref{fig:app}), a HWP flipped the polarization to about 45-degrees from horizontal. It was followed by a quartz wave-plate of about 8 mm in thickness.  The down-conversion into polarization-entangled states uses two thin type-I beta-barium borate (BBO) crystals rotated 90 degrees relative to each other,\cite{KwiatPRA99} which are now quite standard and commercially available (see Table~\ref{tab:parts}). 
The quartz crystal was necessary to temporally shift the pairs coming from the two crystals so that it becomes indistinguishable from which crystal the photons were produced.\cite{RangarajanOE09} 
The tilt of the crystal also provided adjustment of the relative phase between the two polarization possibilities to put the photons in the state of Eq.~\ref{eq:psi+}.

 \begin{table}[h!]
\centering
\caption{Parts list.}
\begin{ruledtabular}
\begin{tabular}{ l l r}
Part & Make and Model & Price each (\$) \\
\hline	
Laser & Power Technology PM50(405)G35 & 1000 \\
BBO crystals & Newlight Photonics PABBO5050-405(I)-HA3 & 1000 \\
Quartz crystal & Newlight Photonics QAR25550-A-AR405 & 429 \\
Thompson prism & Lambda Research CGTS-08 & 770 \\
Wollaston option$^1$ & Optosigma WPPB-06-14SN & 550 \\
Detectors & Excelitas/Alpha SPCM-EDU CD3375 & 1700 \\
Coincidence unit & Red Dog Physics CD48  & 300 \\
Motorized rotation stage & Pacific Laser & 450 \\
\end{tabular}
\end{ruledtabular}
$^1$ Alternative to Thompson prism.
\label{tab:parts}
\end{table}
 
 The photon pairs at nearly the same energy (or wavelength of 805 nm) followed two non-collinear paths. They reached identical detection setups for Alice and Bob that consisted of a HWP followed by a polarization splitter. We tried two different type of splitters. The most inexpensive ones were polarizing beam splitters (about \$250 each), which are widely available. They transmit exclusively horizontally polarized photons and reflect mostly vertically polarized photons. Because of a small contamination of horizontal polarization in the reflection (10-15\%), we followed the splitter with a vertical polarizer, as shown in the figure. This lead to some imbalance in the detection probability so we opted for a different splitter: a Thompson prism. This is an optical  element made of calcite that transmits vertically-polarized photons and reflects at 45 degrees horizontally polarized photons, both with high degree of purity. We already had these prisms, but they are an expensive option (see Table~\ref{tab:parts}). A less expensive option is a Wollaston prism.
 
 Past the splitters we had fiber collimators connected to multimode fibers that channeled the light to 4 single-photon detectors . Before the collimators we had 40-nm band-pass filters. The signals from each detector reached an electronic unit that recorded the detector pulses as they arrived and also recorded ``coincidences'' (pulses from Alice and Bob that arrived within about 50 ns). Our unit was based on a field programmable gate array integrated circuit (Altera model DE2), which  is no longer available, but there is a new option listed in Table~\ref{tab:parts}. The electronic unit was controlled by a laptop via a Matlab program.\cite{colgateurl}
 
 To implement Eve we needed to modify the state of the photon heading toward Bob in a way that mimicked the effect of intercept and resend, as mentioned above. Case I above is easily achieved with a polarization interferometer that uses the same basis as Alice and Bob. This is a 2-path interferometer with polarizing beam splitters such that the light in state $\ket{H}$ goes through one path and light in state $\ket{V}$ goes through the other path. The challenge is to mimic case II: that Bob has 1/2 probability of obtaining the wrong state once Eve picks a basis that is different than the one used by Alice and Bob. This can be achieved also by a polarization interferometer, as mentioned above, but with unequal path-length, as shown in the insert to Fig.~\ref{fig:app}, provided that the difference in path length is greater than the coherence length of the light. 
 
 We explain the effect of the unbalanced interferometer by way of an example: If the photon sent by Alice is in state $\ket{D}$, the horizontal and vertical components (states) are in phase (Eq.~\ref{eq:D}). If the photon meets Bob's polarization splitter in the $DA$ basis, then it goes to the D output with probability 1.  If before reaching Bob's splitter Eve separates the horizontal and vertical components of the photon and delay one relative to the other for a distance longer than their coherence length, then these two components become incoherent. When the paths get recombined, state $\ket{D}$ is not recreated because the light is in an incoherent combination of $\ket{H}$ and $\ket{V}$ states. It is equivalent to the photon being in state $\ket{H}$ half the time and in state $\ket{V}$ the other half of the time. That is, the diagonal state of the photon has been converted from a coherent superposition of horizontal and vertical states into a ``mixed'' state. From a fundamental perspective, the effect of the unbalanced interferometer is to make a measurement in the $HV$ basis in the first polarization splitter and subsequently channel the paths so that regardless of the state into which photon was measured, it continues toward Bob. We can also think that state $\ket{D}$ of Eq.~\ref{eq:D} is a superposition of two indistinguishable possibilities (states $\ket{H}$ and $\ket{V}$). The unbalanced interferometer makes the two possibilities distinguishable because the path length difference is longer than the length of the photon wavepacket (determined by the bandwidth of the filters), and so a timing measurement could in principle reveal whether the photon took the short path (being in state ($\ket{H}$)  or the long path (being in state $\ket{V}$).
 
When the light in this incoherent state reaches Bob's polarization splitter, it will have 1/2 probability of reaching either detector.  Likewise, if we place two half-wave plates (HWP) set to 22.5$\dg$ from the horizontal, one before and one after, then the first HWP transforms respectively the $\ket{D}$ and $\ket{A}$ states to the $\ket{H}$ and $\ket{V}$ states temporarily. After the interferometer the second HWP transforms the states back. This effectively decoheres the two diagonal components, mimicking intercept and resend in the $DA$ basis. 
 
 We started the experiments by implementing the interferometer mentioned above. It had some alignment challenges. However, we discovered that a thick quartz plate (of 8 mm thickness) would do the same due to the short coherence length of the down-converted photons (about 16~$\mu$m, or 54 fs coherence time, when using 40-nm band-pass filters), as had been done before.\cite{NaikPRL00} The quartz plate delays the polarization components along its fast and slow axes by about 207 fs, which is enough to decohere the components of the down-converted photon along the fast and slow axes of the plate. Moreover, by just rotating the fast axis of the quartz plate to the diagonal direction we could decohere the polarization states along the diagonal direction. It was a much simpler alternative for implementing Eve.
 
Finally, by way of advice, there were two challenging aspects of the experiment. The first one was  placing the detectors at the proper locations, at the outputs of the polarization splitters, so that they gave complementary signals. It involved iterations of the signals and detector positions for all the settings of the bases. The second challenge was to get the data acquisition to measure low count rates for the chosen counting interval. Without any attenuation we would get about 200 coincidences per second when Alice and Bob were set to the same basis. We had to reduce this to an average of 1 coincidence count per interval. We accomplished this by first reducing the counting interval to the minimum possible: 0.1~s, determined by our coincidence-circuit unit. We then attenuated the pump beam so that it fed a very weak photon stream to the down-conversion crystal. This entailed a band-pass filter, to make sure only pump photons went through (some diode lasers produce a weak infra-red glow that interferes with the measurements) followed by a HWP and two horizontal polarizers as already mentioned. 
 
\section{Results}\label{sec:results}
 Our data acquisition then consisted of an apparatus that produced pairs of entangled photons, one going to Alice and another one to Bob through Eve. The choice of basis for Alice and Bob  was done by the setting of the HWP that preceded the respective polarization splitter (0$\dg$ for $HV$ and 22.5$\dg$ for $DA$). The base choice for Eve was determined by the angular orientation of the fast axis of the quartz wave plate (0$\dg$ for $HV$ and 45$\dg$ for $DA$). The rotational mounts that we had were motorized and USB-controlled. We note that there are 3-D printed alternatives for the rotational mounts.\cite{ToninelliAOP} Although they were not necessary, they helped automize the acquisition. 
 As mentioned earlier, the photon data was acquired with a dwell time of 0.1 s, and stored automatically on a spreadsheet. We later sorted the data in the spreadsheet. Whenever we did not have one count in only one the two detectors for Alice and Bob, we eliminated the entry, keeping only the cases that could potentially be used to obtain the key.
 
Tables \ref{tab:HVHVHV} and \ref{tab:DADADA} show a sample of our results when Alice, Bob and Eve were set to the same $HV$ and $DA$ bases, respectively.
In our exercise we did 667 trials with equal settings for all three parties and in $92\pm8$\% of cases they agreed. The agreement in the $HV$ basis was greater than in the $DA$ basis. This is likely due to not having perfect state fidelity. We implemented the setup and the experiment within the undergraduate laboratory context, with students (the first 2 authors) doing the alignments and the experiments. The error rate of 8\% was still below the 11\% that is considered acceptable for generating a secret key.\cite{BouchardQ18}

\begin{table}[h!]
\centering
\caption{Sample data when Alice, Bob and Eve are in the $HV$ basis (H=1, V=0).}
\begin{ruledtabular}
\begin{tabular}{c c c c}
Trial & Alice & Bob & Agree?\\
\hline	
1 & 0 & 0 & \checkmark \\
2 & 0 & 0 & \checkmark \\
3 & 0 & 0 & \checkmark \\
4 & 1 & 1 & \checkmark \\
5 & 1 & 1 & \checkmark \\
6 & 0 & 0 & \checkmark \\
7 & 1 & 1 & \checkmark\\
8 & 0 & 0 & \checkmark\\
9 & 0 & 0 & \checkmark  \\
10 & 0 & 0 & \checkmark \\
\end{tabular}
\end{ruledtabular}
\label{tab:HVHVHV}
\end{table}

\begin{table}[h!]
\centering
\caption{Sample data when Alice, Bob and Eve are in the $DA$ basis (D=1, A=0).}
\begin{ruledtabular}
\begin{tabular}{c c c c}
Trial & Alice & Bob & Agree?\\
\hline	
1 & 0 & 0 & \checkmark\\
2 & 1 & 1 & \checkmark\\
3 & 1 & 1 & \checkmark\\
4 & 0 & 1 & $\times$ \\
5 & 0 & 0 & \checkmark\\
6 & 0 & 1 & $\times$ \\
7 & 0 & 0 & \checkmark\\
8 & 0 & 0 & \checkmark\\
9 & 0 & 0 & \checkmark\\
10 & 1 & 1 & \checkmark\\
\end{tabular}
\end{ruledtabular}
\label{tab:DADADA}
\end{table}

We also took data with all the combinations of distinct bases.  Tables \ref{tab:HVHVDA} and \ref{tab:DADAHV} are representative samples of two important cases: when Alice and Bob had the same basis but Eve had a different basis.  As can be seen, they show many more discrepancies between Alice and Bob due to Eve's intervention. Out of 1320 trials of these two cases we found that Alice's and Bob's results disagreed $28\pm4$\% of the time, which is within the expectation of 25\% (assuming that Eve chooses her basis randomly between $HV$ and $DA$). We note that these are the most basic considerations. In a real application other considerations come into place, some of which we describe below.

\begin{table}[h!]
\centering
\caption{Sample data when Alice and Bob are in the $HV$ basis, and Eve in the $DA$ basis (H,D=1, V,A=0).}
\begin{ruledtabular}
\begin{tabular}{c c c c}
Trial & Alice & Bob & Agree? \\
\hline	
1 & 1 & 1 & \checkmark\\
2 & 0 & 0 & \checkmark \\
3 & 1 & 0 & $\times$ \\
4 & 1 & 0 & $\times$\\
5 & 1 & 1 & \checkmark \\
6 & 1 & 1 & \checkmark\\
7 & 1 & 1& \checkmark \\
8 & 0 & 1 & $\times$\\
9 & 0 & 0 & \checkmark\\
10 & 1 & 1 & \checkmark\\
\end{tabular}
\end{ruledtabular}
\label{tab:HVHVDA}
\end{table}

\begin{table}[h!]
\centering
\caption{Sample data when Alice and Bob are in the $DA$ basis, and Eve in the $HV$ basis (H,D=1, V,A=0).}
\begin{ruledtabular}
\begin{tabular}{c c c c}
Trial & Alice & Bob & Agree?\\
\hline	
1 & 0 & 0 & \checkmark\\
2 & 1 & 0 & $\times$ \\
3 & 1 & 1 & \checkmark\\
4 & 1 & 0 & $\times$\\
5 & 1 & 1 & \checkmark\\
6 & 1 & 1 & \checkmark\\
7 & 1 & 1 & \checkmark\\
8 & 0 & 0 & \checkmark\\
9 & 0 & 0 & \checkmark\\
10 & 1 & 0 & $\times$\\
\end{tabular}
\end{ruledtabular}
\label{tab:DADAHV}
\end{table}

\section{Other Measurements: Quantum State Tomography}\label{sec:qst}
This basic setup can be a template for other teaching moments. It can be used in the context of teaching quantum mechanical principles and entanglement. The starting point of this experiment is to create an entangled state. We do this by first aligning the apparatus and measuring coincidence counts in the detectors consistent with the state being prepared.\cite{DehlingerAJP02} An extension of this experiment is to do a Bell test: either a Clauser-Horne-Shimony-Holt (CHSH) test, \cite{CHSH,DehlingerAJP02} or other, such as the Hardy test.\cite{CarlsonAJP05} This involves doing 16 projective measurements on the state of the light plus computation of the terms in the inequality. The state without Eve will violate the Bell inequality, and the one with Eve will not.\cite{EkertPRL91}

\begin{figure}[h!]
\centering
\includegraphics[width=5in]{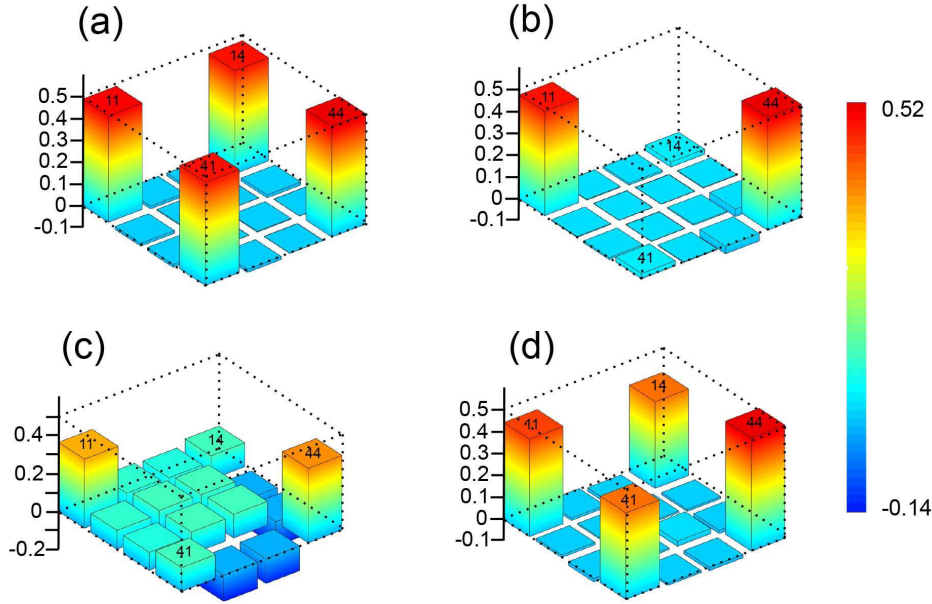}
\caption{Bar graph of the measured real component of the density matrix of the state of the light received by Alice and Bob before their joint measurement. The cases include: (a) when Eve is not present, (b) When Eve decoheres in the $HV$ basis, (c) when Eve decoheres in the $DA$ basis, and (d) when Eve performs partial decoherence.}
\label{fig:qst}
\end{figure}

Another alternative could be to use the experiment in the context of quantum information, and using it to extract the information conveyed by the light. This  involves doing quantum state tomography (QST) on the light. It entails also doing sixteen projective measurements of the state of the two photons to obtain the density matrix of the state.\cite{JamesPRA01,AltepeterAAMP05} To do this we need two additional quarter-wave plates (QWP) so that we can measure right and left handed circular polarization. In Appendix~B we briefly describe how to do the 16 measurements and obtain the density matrix of the light. A simple description of the density matrix has been presented before in this context.\cite{GalAJP10} Briefly, if the basis is $\ket{H}_1\ket{H}_2 =(1\;0\;0\;0)^{\rm T}$, $\ket{H}_1\ket{V}_2 =(0\;1\;0\;0)^{\rm T}$, $\ket{V}_1\ket{H}_2 =(0\;0\;1\;0)^{\rm T}$ and $\ket{V}_1\ket{V}_2 =(0\;0\;0\;1)^{\rm T}$, then the density matrix of the state of Eq.~\ref{eq:psi+} is 
\begin{equation}
\rho_\psi=\ket{\psi}\bra{\psi}
=\frac{1}{2}\left(
    	\begin{array}{cccc}
			1  & 0 & 0 & 1\\
			0 & 0 & 0 & 0\\
			0 & 0 & 0 & 0\\
			1 & 0 & 0 & 1
	\end{array}\right).\label{eq:rpsi}
	\end{equation}
Figure~\ref{fig:qst}(a) shows the results of the quantum state tomography that we took of this state: the bar figure is a standard pictorial way to represent the real component of the matrix elements of the density matrix, with the height of the bar representing the value of the matrix element. As can be seen, the figure faithfully represents the expected form of Eq.~\ref{eq:rpsi}: predominantly 4 bars at the 4 corners of the matrix with values close to 0.5: $\rho_{11}=0.49$, $\rho_{14}=0.47$, $\rho_{41}=0.47$ and $\rho_{44}=0.51$. 

The previous tomography measurement was made without Eve. We followed by adding Eve, an 8-mm-thick quartz plate, as mentioned earlier, with its axis vertical. In this case the optic representing Eve decohered the horizontal and vertical components of Bob's photon, which is equivalent to rendering the light in the mixed state given by 
\begin{equation}
\rho_{\rm w/Eve}=\frac{1}{2}\ket{H}_1\ket{H}_2\bra{H}_1\bra{H}_2+\frac{1}{2}\ket{V}_1\ket{V}_2\bra{V}_1\bra{V}_2
=\frac{1}{2}\left(
    	\begin{array}{cccc}
			1  & 0 & 0 & 0\\
			0 & 0 & 0 & 0\\
			0 & 0 & 0 & 0\\
			0 & 0 & 0 & 1
	\end{array}\right).\label{eq:rfeve}
	\end{equation}
The tomography of this state is shown in Fig.~\ref{fig:qst}(b), which nicely reflects the expectation of Eq.~\ref{eq:rfeve}: 
the bars for the 2 corners have values $\rho_{11}=0.48$, $\rho_{44}=0.52$, whereas the other elements had values in absolute value below 0.05 (e.g., the complex matrix elements were $\rho_{41}=\rho^*_{14}=0.024+0.021i$ and $\rho_{43}=\rho^*_{34}=-0.034-0.036i$).	

By knowing the density matrix one can obtain useful measures of the state of the light.
They include the tangle, a measure of the entanglement ($0=$ not entangled; $1= $  fully entangled). The von Neumann or quantum Shannon entropy is given by\cite{Nielsen} 
\begin{equation}
S=-{\rm Tr}(\rho\log_2\rho)=-\sum_i\lambda_i\log_2\lambda_i,
\end{equation}
where $\lambda_i$ are the eigenvalues of the density matrix (and with $0\log_20\equiv0$). $S$ is a measure of the uncertainty in the information ($0=$ no uncertainty, state is well defined), and  $1= $ full uncertainty with 2 basis states). It is easy to show that $S=0$ for $\rho_\psi$ (due to $\lambda_1=1$, $\lambda_2=\lambda_3=\lambda_4=0$) and $S=1$ for $\rho_{\rm w/Eve}$ (due to $\lambda_1=\lambda_2=1/2$ and $\lambda_3=\lambda_4=0$). The linear entropy is 
another measure of the degree of mixture ($0=$ not mixed; 2/3 = mixture of 2 basis states; $1= $ maximally mixed, or mixture of all basis states). The fidelity is the probability that the measured state is $\psi$. The results for the states just mentioned are given in Table~\ref{tab:qst}. We can see that when Eve is not present we have a highly entangled state, with a high value of tangle and a low value of the von Neumann entropy, reflecting a high certainty in the state of the light and a high probability that it is in state $\psi$. When Eve is present, the tangle is nearly zero and the von Neumann entropy is consistent with 1, revealing full uncertainty in the state of the light (i..e, per density matrix, 48\% chance that it is in state $\ket{H}_1\ket{H}_2$ and 52\% that it is in state $\ket{V}_1\ket{V}_2$). The state is not maximally mixed, although mixed enough to contain no information. We also did tomography for the diagonal setting of Eve (Fig.~\ref{fig:qst}(c)) and confirmed that the tangle is low and the entropy was high, as shown in the third entry of the table. The density matrix for this case, is quite different than the one when Eve is aligned with the vertical. 

\begin{table}[h!]
\centering
\caption{Quantum state tomography of the state of the light for 3 different cases. Full Eve consists of inserting a 0.64-mm thick quartz plate; and for partial Eve we inserted a 1-mm thick quartz plate. In both cases the crystal axis aligned with the vertical.}
\begin{ruledtabular}
\begin{tabular}{l c  c c c}
Case & Tangle & von Neumann Entropy &Linear Entropy & Fidelity \\
\hline	
Without Eve & $0.96\pm0.05$ & $0.1\pm0.2$ & $0.03\pm0.08$ & 0.97 \\
Full Eve aligned vertical & $0.004\pm0.002$ & $1.0\pm0.2$ & $0.66\pm0.03$ & 0.52\\
Full Eve aligned diagonal & $0.001\pm0.002$ & $1.1\pm0.5$ & $0.68\pm0.04$ & 0.50 \\
Partial Eve & $0.68\pm0.08$ & $0.45\pm0.14$ & $0.22\pm0.10$ & 0.88 \\
\end{tabular}
\end{ruledtabular}
\label{tab:qst}
\end{table}

We did an additional test with what we call ``partial Eve''. That is, we used a quartz plate of thickness 0.64 mm. It was a multiple-order waveplate that was available in the lab from a previous project. The thickness of the plate only partially decohered the state. Thus we would expect to produce a state that showed some entanglement and some degree of mixture. Figure~\ref{fig:qst} indeed shows some resemblance of the initial state, and per Table~\ref{tab:qst}, with reduced tangle and a von Neumann entropy in between maximum and minimum uncertainty. Interestingly, in the context of quantum communication, the latter optic mimics the effect of Eve doing intercept and resend of only a fraction of the photons that Alice sends to Bob, obtaining partial information about the communication possibly without being detected. For example, if Eve picked only 1/2 of the photons while alternating equally between $HV$ and $DA$ bases, there would be a probability of 1/8 that Alice and Bob will get errors from the eavesdropping. (If the error rate is low enough this knowledge can be eliminated by privacy amplification, explained in Appendix~A.) A more exhaustive analysis of this situation can be done but we considered it beyond the scope of our project. In Appendix~B we give a brief explanation of the measurements that entail quantum state tomography. 

\section{Conclusions}\label{sec:concl}
In summary, we have developed a laboratory demonstration of the generation and distribution of a cryptographic key (QKD) using single photons. This is the simplest type of QKD but it suffices to convey the spirit of the technology. Undergraduate coauthors were involved in the setup, alignment and data acquisition. The most significant aspect of the demonstration was the use of a thick quartz plate to mimic the intercept and resend action of the eavesdropper. We also have used quantum state tomography to analyze the state of the light and show the effect of the eavesdropper on the state of the light.
 
\begin{acknowledgments}
This work was funded by National Science Foundation grant PHY-1506321. We thank C.H. Holbrow for assistance with the manuscript.

\end{acknowledgments}

\section{Appendix A: Communicating with the Key}
In the communication leading to obtaining the key, Alice sends Bob a stream of photons entangled with the ones she keeps. Next they share over a public channel (that is, through a non-secret communication) the bases that they used for each pair of photons. Then they both keep only the data obtained when both used the same basis ($HV$ or $DA$). This leaves them with a string of binary numbers that is in principle identical for both. Note that Alice, Bob and in principle others, including Eve, know the bases that they used but only Alice and Bob know the outcome of the measurements, which by quantum mechanics are identical. Thus, they each generated the same string of binary numbers without transmitting it explicitly to each other. 

At this point Alice and Bob must determine whether there is an eavesdropper or not. This involves sharing a fraction of their string (via a random sampling) and looking for errors. It is here where information theory comes into action. From a simple analysis, if the errors exceed 25\%, then that indicates that there is an eavesdropper and the string has to be discarded. Additionally, there are always some errors due to instrumental effects or noise, although it is impossible to distinguish them from an eavesdropper attack. If the error rate is below some predetermined percentage, either from instrumental imperfections or Eve under-sampling in her intrusion, Bob and Alice conclude it is ok to proceed and follow steps to correct the errors---a process known as error correction or information reconciliation. Briefly, it involves exchanging the parity of blocks of binary data in a recursive way so that the errors are effectively corrected. Once the errors are corrected, Alice and Bob have the same string. However, the error-correction step may have revealed some information to the eavesdropper, so a second step in the process, called privacy amplification, involves obtaining the final string, the key, from operations between parts of the longer set of numbers. This step insures that the eavesdropper has no access to the key because it involves operations with data she does not have. There is a huge body of technical literature on this subject, but a recent textbook gives simple and clear explanations of these last two steps.\cite{LoeppWooters}

Suppose now that Alice and Bob have a key and wish to communicate. The message is converted to binary consisting of $N$ data bits $D_i$ ($i=1,2\ldots N$). The key also consists of binary bits, so Alice uses $N$ key bits $K_i$. The bits of the encrypted message $E_i$ are obtained by applying the exclusive-OR operation (XOR $=\oplus$) between the data and the key, bit by bit:
\begin{equation}
E_i=D_i\oplus K_i = D_i\cdot K_i'+D_i'\cdot K_i,
\end{equation}
where ``$\cdot$'' and ``$+$'' are the Boolean operations AND and OR, respectively, and the prime on a variable denotes the NOT operation. When Bob receives the string of bits he gets the original message bits by applying the XOR operation between the encrypted bits and the key bits. The result is the message bits. This is possible due to the relation:
\begin{equation}
E_i\oplus K_i=D_i
\end{equation}
It can be left as an exercise for the student to prove the previous expression by applying the following Boolean identities between variables $A$ and $B$:
\begin{eqnarray}
(A+B)'=A'\cdot B'\\
A\cdot A'=0\\
A+A'=1.
\end{eqnarray}
As a subsequent exercise, the student can be asked to verify the relations with a numerical string of message bits and  key bits, and following the XOR truth table given in Table~\ref{tab:xor} (this is the same as binary addition modulo 2).
\begin{table}[h!]
\centering
\caption{Truth table for the XOR Boolean operation for 2 bits.}
\begin{ruledtabular}
\begin{tabular}{ c  c c }
$A$ & $B$ & $A\oplus B$ \\
\hline	
0 & 0 & 0  \\
0 & 1 & 1\\
1 & 0 & 1 \\
1 & 1 & 0 \\
\end{tabular}
\end{ruledtabular}
\label{tab:xor}
\end{table}

\section{Appendix B: Quantum State Tomography of Photon Pairs}\label{sec:qstb}
The state of the light is most accurately represented by the density matrix. In the case of 2 qubits presented here, the density matrix is a $4\times 4$ Hermitian matrix, which  has 4 real diagonal elements $\rho_{ii}$ (with $\sum_i\rho_{ii}=1$) and 12 complex off-diagonal elements with $\rho_{ij}=\rho_{ji}^*$. These plus the total photon count add to a total of 16 unknown quantities. QST can then be done by making 16 projective measurements. 
The matrix elements are obtained from linear combinations of the measurements plus an optimization that enforces Hermitian properties.\cite{JamesPRA01}   

The measurements involved projecting the state of the light through polarization filters (one for Alice and one for Bob). Projection onto a given state, say $\ket{H}$, is achieved by setting polarization optics so that a photon in that state is transmitted without attenuation, and conversely the state orthogonal to it ($\ket{V}$) would be fully blocked.  Projection onto state $\ket{H}$, involves a filter that is a horizontal polarizer. Similar arguments follow with the other linear states $\ket{V}$, $\ket{D}$ and $\ket{A}$. Because projections onto the right and left circular states ($\ket{R}$ and $\ket{L}$, respectively) are needed, we add a quarter-wave plate QWP in front of the polarizer. When detecting the linear states, the QWP should not affect the state. Therefore, it should be aligned along the same direction as the polarizer's transmission axis. To make filters for $\ket{R}$  and $\ket{L}$  we align the fast axis of the QWP along a given direction and followed it with the polarizer with transmission axis aligned 45-degrees relative to the fast axis of the QWP, counter-clockwise for detecting $\ket{R}$ and clockwise for $\ket{L}$, when looking into the incoming beam. The 16 filter settings for Alice and Bob are: $HH$, $HV$, $VV$, $VH$, $RH$, $RV$, $DV$, $DH$, $DR$, $DD$, $RD$, $HD$, $VD$, $VL$, $HL$, $RL$.\cite{JamesPRA01}

In the QKD apparatus we used Thompson prisms as polarization splitters for both Alice and Bob, with the $V$ state transmitted and $H$ state deflected. To avoid inconsistencies due to differing detector efficiencies we did all the measurements with the recordings of the straight output of the two Thompson prisms.  Because the prisms were fixed, the HWP before each prism had to be adjusted appropriately. A summary of settings is given in Table~\ref{tab:polariz}. 
\begin{table}[h!]
\centering
\caption{Angles of quarter and HWPs for produce the appropriate filtering action when the two waveplates are followed by a vertical polarizer. Angles are in degrees, with positive being counter-clockwise looking into the beam, and zero being the vertical direction.}
\begin{ruledtabular}
\begin{tabular}{c r r}
State & Qwp & Hwp \\
\hline	
$H$ & 90 & 45 \\
$V$ & 0 & 0 \\
$D$ & $-45$ & $-22.5$ \\
$A$ & $+45$ & $+22.5$ \\
$R$ & 0 & $+45$ \\
$L$ & 0 & $-45$ \\
\end{tabular}
\end{ruledtabular}
\label{tab:polariz}
\end{table}

Once the coincidence readings for each of the 16 setting are obtained, they are fed into the QST program that computes the density matrix and quantum measures. We used a program based on an online resource from Paul Kwiat's research group.\cite{KwiatQST} We can also provide our own version of the program upon request. 
\end{document}